# Forward Private Searchable Symmetric Encryption with Optimized I/O Efficiency

Xiangfu Song, Changyu Dong, Dandan Yuan, Qiuliang Xu and Minghao Zhao

**Abstract**—Recently, several practical attacks raised serious concerns over the security of searchable encryption. The attacks have brought emphasis on forward privacy, which is the key concept behind solutions to the adaptive leakage-exploiting attacks, and will very likely to become mandatory in the design of new searchable encryption schemes. For a long time, forward privacy implies inefficiency and thus most existing searchable encryption schemes do not support it. Very recently, Bost (CCS 2016) showed that forward privacy can be obtained without inducing a large communication overhead. However, Bost's scheme is constructed with a relatively inefficient public key cryptographic primitive, and has a poor I/O performance. Both of the deficiencies significantly hinder the practical efficiency of the scheme, and prevent it from scaling to large data settings. To address the problems, we first present FAST, which achieves forward privacy and the same communication efficiency as Bost's scheme, but uses only symmetric cryptographic primitives. We then present FASTIO, which retains all good properties of FAST, and further improves I/O efficiency. We implemented the two schemes and compared their performance with Bost's scheme. The experiment results show that both our schemes are highly efficient, and FASTIO achieves a much better scalability due to its optimized I/O.

**Index Terms**—searchable encryption, symmetric primitives, forward privacy, I/O efficiency.

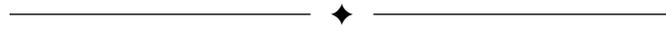

## 1 INTRODUCTION

SEARCHABLE encryption is perhaps one of the most intensively studied cryptographic primitives. The need for searchable encryption comes from the surge of Storage-as-a-service, a service delivery model in which the clients store their data on remote servers that are managed by external service providers (e.g. Amazon S3, Microsoft Azure Storage, Google Cloud Storage etc.). When data storage is outsourced, data privacy becomes a primary concern because the service providers may not always be trusted. While conventional encryption can be used to protect the outsourced data, it does not allow effective retrieval of data by searching on encrypted data. Searchable encryption was proposed to solve this problem. In general, searchable encryption schemes work by generating an encrypted index, which will be outsourced to the cloud service provider along with the encrypted data. Later the client can generate search tokens that encrypt certain keywords and the server can perform a search algorithm using the tokens and the encrypted index to find matches. In this way, even though data is stored on and searching is handled by an untrusted server, the privacy of the data can still be preserved.

Recently, there has been a major concern over the security of searchable encryption schemes. Since 2012, several attacks [1], [2], [3] have been devised that allow an untrusted server to recover the keywords in the client's search tokens, and in consequence, to learn a significant amount of information about the outsourced encrypted data. These attacks are normally performed by utilizing information leaked in the searching and updating phase, and they are pervasive because the information leakage exists inevitably in any searchable encryption scheme. In the most recent work by Zhang et al [3], the authors showed a simple yet effective adaptive attack that can fully reveal the client's queries by injecting only a small number (usually less than 100) of files to the encrypted data store. The result is devastating: not only it enables the server to learn partial information about the encrypted data, the recovered keywords can also help the server in other statistical attacks. The essential idea of the attacks is that the server first crafts a set of files (each contains certain keywords), then sends the files to the client and tricks the client into encrypting them. After the client has encrypted and uploaded the injected files, the server can use the tokens previously submitted by the clients to search on the injected files. By knowing which keywords are in each injected file and observing which files matches the token, the server can deduce easily which keyword is encrypted in the token. The attacks show that even seemingly harmless small leakage can be exploited, and highlight the importance of *forward privacy*, which requires that a newly inserted file cannot be linked in anyway with previous search queries. With forward privacy, the attacks can be prevented.

While forward privacy is not a new concept, most searchable encryption schemes do not support it. Until recently, forward private searchable encryption implies prohibitively high communication cost. There is a trivial way to achieve forward privacy: when a new file needs to be added, the client downloads all encrypted files from the server, re-indexes them with the new file and re-encrypts everything with new keys. This is obviously impractical. There were also non-trivial solutions using an oblivious data storage, e.g. Oblivious RAM (ORAM)[4], [5] or Distributed Oblivious Data structure (DOD) [6], which hides search patterns and access patterns from the server. In this approach, the server cannot observe which files matches the token, therefore cannot learn anything about the query. Nevertheless, the communication

- X. Song, D. Yuan and Q. Xu are with the School of Computer Science and Technoloy, Shandong University, 250101 Jinan, China.
  E-mail: bintasong@gmail.com, dandanyuan.sdu@gmail.com, xql@sdu.edu.cn
- C. Dong is with the School of Computing, Newcastle University, NE4 5TG Newcastle Upon Tyne, U.K.
  E-mail: changyu.dong@newcastle.ac.uk
- M. Zhao is with the School of Software, Tsinghua University, 100084 Beijing, China.
  E-mail: mh-zhao17@mails.tsinghua.edu.cn

cost is still too high unless replying on strong assumptions like multiple non-colluding servers. One notable work by Chang and Mitzenmacher [7] achieves forward privacy without using an oblivious data storage. However, in this scheme the search query size grows linearly in the number of updates, which means after the system has run for a period of time, the communication cost for the search operation will become unacceptably high due to the ever-growing query size.

In CCS 2016, Bost [8] proposed an efficient forward secure searchable encryption scheme called Sophos that achieves a low communication cost. In this scheme, the client keeps a state for each keyword and evolves the state when updating a file on the server that contains this keyword. The state is used as an input when generating the encrypted index for the updated file, which means the previous token is outdated and cannot be used to match the new index.The state is kept secret from the server util the client performs a search, at which point a constant size token is generated that contains the current state. The token enables the server to recover all previous states from the current state. By knowing all states, the server can search through all updates and find matching files. Sophos uses trapdoor permutation, a public key primitive, for evolving the state so that the server, who does not know the private key, cannot predict future states. Thus when new updates come in, the server learns nothing and forward privacy is achieved.

Although Sophos is more efficient than previous forward private schemes, there are still two significant deficiencies in the design. Firstly, the scheme is based on public key cryptography. It has been showed clearly in Bost's paper [8] that if the data is updated frequently, the public key operations become a major performance bottleneck. It would be ideal if only symmetric cryptographic primitives are used. Secondly, the search operation is not I/O efficient. Recently it has been shown that I/O has become a major bottleneck that prevents searchable encryption from scaling up to large data [9], [10]. In Sophos, the locality, a measure of I/O efficiency, of the search operation increases linearly in the number of updates. The performance of the search operation is inversely proportional to the locality, which means search performance will degrade inevitably and eventually become unacceptably slow after a long run. The author acknowledged this problem but also argued that the poor locality is a necessary price paid for forward privacy.

## 1.1 Our Contributions

Our investigation started from two questions: 1) can we construct a communication-efficient forward private searchable encryption scheme using only symmetric key primitives? 2) to what extent can we optimize the I/O efficiency under the security constraint imposed by forward privacy? We present positive answers to these two questions by constructing two new searchable encryption schemes FAST and FASTIO.

- FAST (**F**orward priv**A**te **S**ymmetric searchable encryp**T**ion) is our first attempt to construct a forward private searchable encryption scheme using only symmetric key primitives. The design of FAST follows the state-based approach used by Sophos [8], but with a very different idea. In Sophos, the states are related by a trapdoor permutation. Since the trapdoor permutation is a public key primitive, by giving the server only the public key, the server can generate states backwards, but not forwards. This strategy however is not valid when using symmetric key primitives. The idea in FAST is inspired by singly linked list, a classic data structure. Each update is like a node in a singly linked list and contains a randomly generated ephemeral key (encrypted). The key is like the pointer in a list node that points backwards to the previous node. In FAST, a new state is generated by encrypting the previous state using the ephemeral key. Given the current key and the current state, the server can compute the previous state by decryption. The decrypted state will help the server to get the previous update and then the previous key. Iteratively, the server can find all previous states as well as all previous updates. However, since the key is randomly generated on-the-fly by the client, by no means could the server predict the future keys/states. Thus forward privacy is achieved.

- FASTIO, where IO in the name stands for **I/O O**ptimized. While FAST is entirely based on symmetric key cryptography which makes it computationally more efficient, it also results in bad I/O efficiency. The I/O overhead comes from 1) the ephemeral keys which is required to be stored on the server, and 2) the fact that the server is forced to read the index in a very non-local manner in order to achieve the forward privacy property. At first glance, the I/O inefficiency seems to be the necessary tradeoff in exchange for other good properties. However, this is not exactly the case. It was discovered after a further investigation that the information leaked in previous queries can actually be utilized to improve I/O efficiency while keeping all other properties intact. The key idea in FASTIO is that after each search query, we can let the server store the result. Since the search result has already been revealed to the server, saving it will not leak additional information to the server. By doing so, when the next search query comes in for the same keyword, the server does not need to search again from scratch. Instead, the server only needs to search through the new updates since the last search and combine the new result with the previous search query result. In this way, the server needs only the latest state and does not need to store ephemeral keys anymore. In addition, the non-local reads are also minimized because the previous search result can be read continuously and non-local reads are only needed for searching through the new updates. While I/O overhead is lowered, all operations can still be based on symmetric key cryptography, and forward privacy can still be achieved.

We implemented the two schemes and conducted experiments to measure their performance. The results confirm that our schemes are more efficient in terms of computation than Sophos, which is the most efficient forward private searchable encryption scheme to date. The results also confirm that FASTIO has a much better I/O efficiency, which makes its much more scalable.

## 2 RELATED WORK

**Searchable encryption** The functionality of secure search can be realized by using generic cryptographic primitives such as Fully Homomorphic Encryption (FHE) [11], Oblivious RAM (ORAM)[12], [13] and Secure Multi-Party Computation(SMPC) [14], [15]. With the generic primitives, it is possible to perform secure search with minimum information leakage. However, searchable encryption based on generic primitives are inefficient and are not suitable to used in practice. Research in practical searchable encryption started from the seminal paper by Song et al. [16], and the goal is to achieve practical efficiency and scalability with the price of accepting a small leakage.

One branch in this research area considers the static setting, in which the client encrypts the data at the setup phase, and does not update the ciphertexts on the server side afterwards. In this setting, Curtmola et al. [17] gave the first reversed-index based scheme with sub-linear efficiency. Chase and Kamara [18] also proposed a similar scheme but with higher storage complexity. Cash et al. [9] proposed a scheme that support boolean queries. Static searchable encryption, although can achieve sub-linear efficiency, are not suitable in many application scenarios where updates are required.

Dynamic searchable encryption enables the client to add and delete data items after the dataset is outsourced to the cloud. Actually, the work of Song et al. [16] supports dynamically update, but the search and update require sequential scan performed on the full-text. Kamara et al. [19] proposed a dynamic searchable symmetric encryption (DSSE) scheme as an improvement of [17]. Their scheme achieves sub-linear search time but has a high complexity for the *add* and *delete* operations. The scheme also leaks much more information to the server than [17]. Dong et al. [20] proposed a dynamic searchable encryption scheme that supports multiple users. However the scheme is based on public key cryptography. Kamara and Papamanthou [21] proposed a parallel DSSE scheme, which leaks less information but require multiple interactions between the server and client. Sun et al. [22] proposed a multi-client version of DSSE scheme that support boolean query. Hahn and Kerschbaum [23] proposed a scheme that only leaks the search pattern meanwhile achieves linear searching complexity. Naveed et al. [24] proposed a new primitive called blind-storage, and based on it, they constructed a DSSE scheme. Instead of processing and calculating on the data stored on it, in this scheme, the server just acts as a transmission and storage entity. It achieves less information leakage but requires multiple rounds of interaction between the client and server. Gajek [25] proposed a DSSE scheme based on constrained functional encryption (CFE), which encodes the index as an encrypted binary tree and constructed an inner product functionality to traverse the encrypted tree. Another DSSE scheme with multi-keyword ranked search was given by Xia et al. [26].

**Forward Privacy** Roughly, forward privacy requires that a newly inserted file cannot be matched with previous search queries. This notion is not applicable to static searchable encryption schemes because they do not support update. Most dynamic searchable encryption schemes do not support forward privacy. To date, most existing forward secure dynamic searchable encryption schemes are constructed based on ORAM-like structure. Among them, the conception of forward privacy was firstly precisely stated by Stefanov et al. [4]. They proposed a forward private DSSE scheme based on hierarchical ORAM. Likewise, Garg et al. [5] proposed the first efficient round-optimal oblivious RAM(ORAM) scheme. As an application of their ORAM scheme, they constructed a DSSE scheme with sub-linear search efficiency meanwhile concealing the search pattern. The problem of this approach is the high communication cost that is resulted from using ORAM like structures. In the scheme of Hoang et al.'s [6], they use a distributed oblivious data structure that is a distributed encrypted incidence matrix created on two non-colluding servers. There are only a few forward private searchable encryption schemes that do not use ORAM. In the scheme proposed by Chang and Mitzenmacher [7], forward privacy was achieved by changing search token after each update. The tokens are unrelated thus to perform a search the client needs to submit all previous tokens in the query. The most efficient forward private searchable encryption scheme was proposed by Bost [8], in which forward privacy is achieved by using a one-way trapdoor permutation to evolve a state stored at the client side.

**I/O Efficiency of Searchable Encryption** Focusing on the practical performance of searchable encryption, Cash et al. [9] made a compromise in security and efficiency and proposed a scheme with good I/O efficiency based on T-Set (a data structure of keyword-entry tuple). Their method was extended by Cash et al. [27]. The new scheme has a better performance in both searching and updating phase. Cash and Tessaro [10] investigated the I/O efficiency of searchable encryption and proved that it is impossible to achieve optimal server storage, read efficiency and locality at the same time due to the security requirement of searchable encryption. Following the work of Cash and Tessaro, a tight lower bound of storage locality was given by Asharov et al. [28]. A tunable SSE scheme was given by Demertzis et al. [29] in which the locality can be tuned. Recently, Demertzis et al. [30] proposed a linear-space searchable encryption scheme with constant locality and sublogarithmic read efficiency. All these schemes are mainly for the static setting and do not support dynamic update. Miers et al. [31] came up with a DSSE scheme by adapting ORAM as oblivious update index(OUI) as an intermediate cache and push a bucket of document identifiers from OUI to append-only storage only when that bucket is filled. Their scheme has a higher IO efficiency compared to original ORAM scheme and achieves a better locality through the bucket mechanism, but leaks more information and does not support forward privacy.

## 3 PRELIMINARIES

### 3.1 Notations

We use $r \xleftarrow{\$} X$ to denote that an element $r$ is sampled uniformly at random from a set $X$, $\{0,1\}^l$ to denote the set of all $l$-bit strings, $\{0,1\}^*$ to denote the set of arbitrary length strings, and $a||b$ to denote the concatenation of two strings $a$ and $b$. Let $\lambda$ be a security parameter, we say a function $\nu : \mathbb{N} \to \mathbb{R}$ is negligible in $\lambda$ if for every positive polynomial $p$, $\nu(\lambda) < 1/p(\lambda)$ for sufficiently large $\lambda$. We use **poly($\lambda$)** and **negl($\lambda$)** to represent the unspecified polynomial and negligible functions in $\lambda$, respectively. For a set $S$, we use $|S|$ to denote $S$'s cardinality. For a string $a$, we use $|a|$ to denote $a$'s bit length.

### 3.2 Searchable Encryption

A *dynamic searchable symmetric encryption* (DSSE) scheme enables a client to outsource its data to an untrusted server in an encrypted form. Later the server can execute search or update queries on the encrypted data. We only consider the dynamic setting in this paper because forward privacy is not applicable to the static setting.

Let a database $\mathsf{DB} = \{(\mathsf{ind}_i, \mathsf{W}_i)\}_{i=1}^{D}$ be a $D$-vector of identifier/keyword-set pairs, where $\mathsf{ind}_i \in \{0,1\}^l$ is a document identifier and $\mathsf{W}_i \subseteq \mathcal{P}(\{0,1\}^*)$. The universe of keywords of the database $\mathsf{DB}$ is $\mathsf{W} = \cup_{i=1}^{D} \mathsf{W}_i$. We use $N = \sum_{i=1}^{D} |\mathsf{W}_i|$ to denote the number of document/keyword pairs. We use $\mathsf{DB}(w) = \{\mathsf{ind}_i | w \in \mathsf{W}_i\}$ to denote the set of documents that contain the keyword $w$. A DSSE scheme $\Pi = \{\mathsf{Setup}, \mathsf{Search}, \mathsf{Update}\}$ consists of three protocols ran by the client and the server:

- $((K, \sigma); \mathsf{EDB}) \leftarrow \mathsf{Setup}(\lambda, \mathsf{DB}; \bot)$: It takes a security parameter $\lambda$ and a database $\mathsf{DB}$ as inputs and outputs $(K, \sigma)$ to the client and EDB to the server, where $K$ is a secret key, $\sigma$ is the client's state, and EDB is the encrypted database.

- $((\sigma', \mathsf{DB}(w)); \mathsf{EDB}') \leftarrow \mathsf{Search}(K, \sigma, w; \mathsf{EDB})$: The client's input consists of its secret key $K$, the state $\sigma$ and a keyword $w$, the server's input is the encrypted database EDB. The client's output include a possibly updated state $\sigma'$ and $\mathsf{DB}(w)$, i.e. the set of the identifiers of the documents that contain the keyword $w$. The server's output is the possibly updated encrypted database $\mathsf{EDB}'$.
- $(\sigma'; \mathsf{EDB}') \leftarrow \mathsf{Update}(K, \sigma, ind, w, op; \mathsf{EDB})$: The client's input is the secret key $K$, the state $\sigma$, a document identifier $ind$, a keyword $w$ and an operation type $op$. The server's input is EDB. The operation $op$ is taken from the set $\{\mathrm{add}, \mathrm{del}\}$, which means the client wants to add or delete a document/keyword pair. The client updates its state and the server updates EDB as requested by the client.

### 3.3 Security Definition

All existing searchable encryptions schemes leak more or less some information to the server, as a tradeoff to gain efficiency. Thus, the security of searchable encryption is defined in the sense that no more information is leaked than allowed. This is captured by using the simulation paradigm and providing the simulator a set of predefined leakage functions $\mathcal{L} = \{\mathcal{L}_{\mathrm{Setup}}, \mathcal{L}_{\mathrm{Search}}, \mathcal{L}_{\mathrm{Update}}\}$.

*Definition 1 (Adaptively Secure Searchable Encryption).* Let $\Pi = (\mathsf{Setup}, \mathsf{Search}, \mathsf{Update})$ be a searchable encryption scheme, $\mathcal{A}$ be an adversary, $\mathcal{S}$ be a simulator parameterized with leakage function $\mathcal{L} = \{\mathcal{L}_{\mathrm{Setup}}, \mathcal{L}_{\mathrm{Search}}, \mathcal{L}_{\mathrm{Update}}\}$. We define the following two probabilistic experiments:

- **Real**$_\mathcal{A}^\Pi(\lambda)$: $\mathcal{A}$ chooses a database DB, the experiment runs $\mathsf{Setup}(\lambda, \mathsf{DB}; \bot)$ and returns EDB to $\mathcal{A}$. Then, the adversary adaptively chooses queries $q_i$. If $q_i$ is a search query then the experiment answers the query by running $((\sigma_{i+1}, \mathsf{DB}_i(w_i)), \mathsf{EDB}_{i+1})) \leftarrow \mathsf{Search}(K, \sigma_i, q_i; \mathsf{EDB}_i)$. If $q_i$ is an update query, then the experiment answers the query by running $(\sigma_{i+1}, \mathsf{EDB}_{i+1}) \leftarrow \mathsf{Update}(K, \sigma_i, q_i; \mathsf{EDB}_i)$. Finally, the adversary $\mathcal{A}$ outputs a bit $b \in \{0, 1\}$.
- **Ideal**$_{\mathcal{A}, \mathcal{S}}^\Pi(\lambda)$: $\mathcal{A}$ chooses a database DB. Given the leakage function $\mathcal{L}_{\mathrm{Setup}}(\mathsf{DB})$, the simulator $\mathcal{S}$ generates an encrypted database $\mathsf{EDB} \leftarrow \mathcal{S}(\mathcal{L}_{\mathrm{Setup}}(\mathsf{DB}))$ and returns it to $\mathcal{A}$. Then, the adversary adaptively chooses queries $q_i$. If $q_i$ is a search query, the simulator answers the query by running $\mathcal{S}(\mathcal{L}_{\mathrm{Search}}(q_i))$. If $q_i$ is an update query, the simulator answers the query by running $\mathcal{S}(\mathcal{L}_{\mathrm{Update}}(q_i))$. Finally, the adversary $\mathcal{A}$ outputs a bit $b \in \{0, 1\}$.

We say $\Pi$ is an $\mathcal{L}$-adaptively-secure searchable encryption scheme if for any probabilistic, polynomial-time (PPT) adversary $\mathcal{A}$, there exists a PPT simulator $\mathcal{S}$ such that:

$$|\Pr(\mathbf{Real}_\mathcal{A}^\Pi(\lambda) = 1) - \Pr(\mathbf{Ideal}_{\mathcal{A},\mathcal{S}}^\Pi(\lambda) = 1)| \leq \mathbf{negl}(\lambda)$$

### 3.4 Leakage Functions and Forward Privacy

A good searchable encryption scheme should reveal as little as possible information. The leakage is captured by the leakage functions. Usually we require $\mathcal{L}_{\mathrm{Setup}} = (D, N)$, i.e. the Setup protocol only leaks the size of DB and the number of document/keyword pairs. For the Search protocol, we usually require $\mathcal{L}_{\mathrm{Search}} = (\mathsf{ap}, \mathsf{qp})$, where ap is the *access pattern* and qp is the *query pattern*. More formally, $\mathcal{L}_{\mathrm{Search}}$ keeps a history

$\mathsf{Hist} = \{(\mathsf{DB}_i, q_i)\}_{i=0}^Q$, which contains all queries $q_i$ so far and the snapshot of the database $\mathsf{DB}_i$ corresponding to $q_i$. The access pattern is defined as $\mathsf{ap}(\mathsf{Hist}) = (t_1, \ldots, t_Q)$ where $t_i = (i, \mathsf{DB}_i(w_i))$ if $q_i = w_i$ is a search query, or $t_i = (i, op_i, ind_i)$ if $q_i = (ind_i, w_i, op_i)$ is an update query. The query pattern is defined as $\mathsf{qp}(w) = \{j | q_j \text{ contains } w \text{ for each } q_j \text{ in } \mathsf{Hist}\}$.

**Forward privacy** is a strong property regarding the leakage of update operation in DSSE. Intuitively, it requires that an update query leaks no information about the updated keyword. The formal definition is as the following:

*Definition 2 (Forward Privacy).* An $\mathcal{L}$-adaptively-secure SSE scheme is forward private if for an update query $q_i = (ind_i, w_i, op_i)$, the update leakage function $\mathcal{L}_{\mathrm{Update}}(q_i) = (i, op_i, ind_i)$.

For the convenience of proof, in our definition the update query contains only one keyword. This definition is actually equivalent to the one in [8], where the update query contains multiple keywords.

### 3.5 Searchable Encryption and I/O Efficiency

Scalability is a key consideration when designing a searchable encryption scheme because the target application is data storage outsourcing, which usually implies that there is a large amount of data. Recent searchable encryption schemes are often very efficient in terms of computation so one consequence is that I/O efficiency becomes the bottleneck for scaling [9], [10], [31], [29].

The I/O efficiency of searchable encryption is characterized by three factors: server side index size, the locality and the read efficiency. The first is self-explanatory. The locality is the number of non-contiguous reads the server must perform when answering a search query. High locality means more random access I/O requests and a longer latency due to the inter-requests latency. The read efficiency is quantified by the amount of unnecessary or irrelevant data read. More formally, we have:

*Definition 3 (Locality).* Let $\Pi$ be a searchable encryption scheme, $K$ be a secret key, $\sigma$ be the client's state and EDB be a database encrypted by $\Pi$ under $K$ with regard to $\sigma$. For every $w \in \mathsf{W}$, we say the locality is $r$ for the search query $\mathsf{Search}(K, \sigma, w; \mathsf{EDB})$ if the server can answer the query by reading at most $r$ intervals from EDB.

*Definition 4 (Read Efficiency).* Let $\Pi$ be a searchable encryption scheme, $K$ be a secret key, $\sigma$ be the client's state and EDB be a database encrypted by $\Pi$ under $K$ with regard to $\sigma$. For every $w \in \mathsf{W}$, we say the read efficiency is $c$ for the search query $\mathsf{Search}(K, \sigma, w; \mathsf{EDB})$ if the server can answer the query by reading at most $c \cdot (\sum_{s_i \in \mathsf{DB}(w)} |s_i|)$ bits.

Cash et al. [10] noticed that the security property has a negative effect on I/O efficiency and pointed out that, for sake of security, a searchable encryption scheme must either extend the encrypted index to an overwhelming large size, or perform searching in a non-local way (e.g. in a random manner), or read much more bits than actually needed. Afterwards Bost [8] additionally observed that locality and forward privacy are two irreconcilable notions. On the one hand, forward privacy requires that in an update operation for a keyword $w$, the locations in the EDB that are modified should be unrelated to the locations that already known to match $w$. On the other hand, to reduce locality, it is necessary to organize entries in EDB relating to $w$ together so that they can be read continuously. Bost concluded, which is not entirely true as we will show later, that under the constraint of forward privacy, no locality optimization is possible for keyword $w$ unless large modification is done to the encrypted database, either during searches or updates.

## 4 FAST: FORWARD PRIVATE SYMMETRIC SEARCHABLE ENCRYPTION

In this section, we present FAST, our first forward private searchable encryption scheme. The main design goal of this scheme is to eliminate public key operations while retaining optimal communication complexity.

Recall that a forward private searchable encryption scheme can be based on ORAM. While ORAM can be purely symmetric key based [12], [13], [32], [33], [34], it has a large communication overhead. Therefore, we turned to the state-based approach that was used in a few existing schemes [7], [8]. The main challenge of this approach is how to allow the server to re-generate all previous states (so that communication cost can be reduced), but not later states (so that forward privacy can be achieved). Sophos [8] solved this problem by using a public key primitive (trapdoor permutation), so that the client can evolve the state forward using a private key and the server can re-generate the whole history of the states using the public key. However, there exists a huge gap in adopting this strategy with symmetric key cryptography. Symmetric key means there is only one key and once it is released to the server, the server can compute future states and there is no forward privacy any more. This is why in [7], the client has to keep the key secret and generate the history of states locally rather than giving the key to the server and letting the server to generate the history. However since the query contains the full history of the state, this results in a huge expansion in the token size and a large communication overhead.

The idea of FAST is that rather than pseudorandomly generating all states using a fixed key, the client chooses a random ephemeral key every time to evolve the state. The current state is essentially the encryption of the previous state under the ephemeral key. The ephemeral key is then stored on the server side, encrypted using the current state. The server can only get the ephemeral key if it knows the current state. Then by decrypting the current state with the ephemeral key, the server can derive the previous state. Iteratively, the server obtains all states, which enable it to search. Forward privacy is guaranteed since the server cannot infer unknown states from currently known states and keys, and the search token size can be made constant because the client only needs to give the latest state to the server.

Fig. 1 shows the protocols of the scheme. We also depict in Fig. 2 how the states evolve forward and backward in the update and search protocols. In the Setup protocol, the client generates $k_s$ and $\Sigma$, where $k_s$ is a $\lambda$-bit long-term key that will be used to encrypt keywords and $\Sigma$ is an empty map that will be used to store the states on the client side. The long-term key $k_s$ is necessary to prevent the server from generating tokens by itself. The server generates **T**, which is an empty map to be used to store the encrypted index. In the Update and the Search protocols, $F$ is a pseudorandom function, $P$ is a pseudorandom permutation (and $P^{-1}$ is the inverse permutation), and $h, H_1, H_2$ are secure hash functions with appropriate output lengths. When updating a file that contains a keyword $w$ and whose identifier is $ind$, the client needs to retrieve the previous state $st_c$ from the local state store $\Sigma$ (line 5-8), then generates a random ephemeral key $k_{c+1}$ and evolves the state forward to the current state $st_{c+1}$ using the pseudorandom permutation (line 9-11). The ephemeral key $k_{c+1}$ is not stored on the client side, but is embedded in the encrypted index entry $e$ that will be stored on the server side (line 12). The client also generates a reference $u$ from the current state and the keyword (line 13). The pair $(u, e)$ is sent to the server and the server updates its map **T** accordingly (line 15). To search a keyword $w$, the client retrieves the current state from $\Sigma$ and sends a search token that contains the encrypted keyword and the current state to the server. Given the current state, the server needs to generate all previous states $st_i$ and find the corresponding update sequence. In the for loop of the algorithm, the server can recover the ephemeral key $k_i$ (line 26), which can later be used to recover the previous state (line 36). Because both "add" and "del" are allowed update operations, the server needs to make sure not including deleted files in the result set. To do so, the server maintains a set $\Delta$ that contains deleted file identifiers during the search process. When the server sees a delete update, the server puts the file identifier $ind$ into $\Delta$. When the server sees an add update and the file identifier $ind$ is in $\Delta$, the server removes $ind$ from $\Delta$. The idea is that the server searches backwards with regards to the update sequence, therefore $ind \in \Delta$ means the added file later was deleted. So the two operations cancelled, then $ind$ should be removed from $\Delta$ and should not be added to the result set. If the server sees an add update and the file identifier is not in $\Delta$ then the file was not deleted and the identifier is added to the result set.

**Complexity Analysis** On the client side, the computational complexity for update and search is $O(1)$. The client stores a $\lambda$-bit key and a map $\Sigma$ whose size is $O(|\mathsf{W}|)$, where $|\mathsf{W}|$ is the total number of keywords. On the server side, the computational complexity for update is $O(1)$, and for search is $O(c)$ where $c$ is the total number of updates (add + del) that contains the keyword being searched since the initialization of the system. The server stores a map **T** whose size is $O(N)$ where $N$ is the total number of document/keyword pairs. The communication complexity is $O(1)$ for update. For search, the communication complexity is $O(1)$ for the query and $O(|\mathsf{DB}(w)|)$ for the result set, which contains $|\mathsf{DB}(w)|$ matching files. Complexity-wise, FAST is optimal (given the security constraints).

**Security Analysis** FAST satisfies the adaptive security definition (Definition 1). The search query leaks only the access pattern and query pattern, which is standard in searchable encryption. The update query leaks no information about the keyword, thus satisfies forward privacy. We have the following theorem regarding the security of FAST:

***Theorem 1.*** Let $F$ be a pseudorandom function, $P$ be a pseudorandom permutation. Let $H_1$ and $H_2$ be two hash functions modeled as random oracles. Define leakage $\mathcal{L} = (\mathcal{L}_{\text{Setup}}, \mathcal{L}_{\text{Search}}, \mathcal{L}_{\text{Update}})$ as
$$\mathcal{L}_{\text{Setup}} = \bot$$
$$\mathcal{L}_{\text{Search}}(w) = (\mathsf{ap}(w), \mathsf{qp}(w))$$
$$\mathcal{L}_{\text{Update}}(i, \mathsf{op}_i, w, \mathsf{ind}_i) = (i, \mathsf{op}_i, \mathsf{ind}_i)$$
Then FAST is a $\mathcal{L}$-adaptively-secure dynamic SSE with forward privacy.

The proof can be found in Appendix A.

## 5 FASTIO: I/O OPTIMIZED SCHEME

In this section, we present FASTIO, our second forward private searchable encryption scheme. FASTIO retains all good properties

## Figure 1: Pseudocode of Protocols in FAST

**Setup**$(\lambda, \bot; \bot)$

*Client:*
1: $k_s \xleftarrow{\$} \{0,1\}^\lambda$
2: $\Sigma \leftarrow$ empty map

*Server:*
3: $\mathbf{T} \leftarrow$ empty map

**Update**$(k_s, \Sigma, ind, w, op; \mathbf{T})$

*Client:*
4: $t_w \leftarrow F(k_s, h(w))$
5: $(st_c, c) \leftarrow \Sigma[w]$
6: **if** $(st_c, c) = \bot$ **then**
7: $\quad st_0 \xleftarrow{\$} \{0,1\}^\lambda, c \leftarrow 0$
8: **end if**
9: $k_{c+1} \xleftarrow{\$} \{0,1\}^\lambda$
10: $st_{c+1} \leftarrow P(k_{c+1}, st_c)$
11: $\Sigma[w] \leftarrow (st_{c+1}, c+1)$
12: $e \leftarrow (ind||op||k_{c+1}) \oplus H_2(t_w||st_{c+1})$
13: $u \leftarrow H_1(t_w||st_{c+1})$
14: send $(u, e)$ to server

*Server:*
15: $\mathbf{T}[u] = e$

**Search**$(k_s, \Sigma, w; \mathbf{T})$

*Client:*
16: $t_w \leftarrow F(k_s, h(w))$
17: $(st_c, c) \leftarrow \Sigma[w]$
18: **if** $(st_c, c) = \bot$ **then**
19: $\quad$ return $\emptyset$
20: **end if**
21: send $(t_w, st_c, c)$ to Server

*Server:*
22: $\mathbf{ID}, \Delta \leftarrow \emptyset$
23: **for** $i = c$ **to** $1$ **do**
24: $\quad u \leftarrow H_1(t_w||st_i)$
25: $\quad e \leftarrow T[u]$
26: $\quad (ind, op, k_i) \leftarrow e \oplus H_2(t_w||st_i)$
27: $\quad$ **if** $op$ = "del" **then**
28: $\quad\quad \Delta \leftarrow \Delta \cup \{ind\}$
29: $\quad$ **else if** $op$ = "add" **then**
30: $\quad\quad$ **if** $ind \in \Delta$ **then**
31: $\quad\quad\quad \Delta \leftarrow \Delta - ind$
32: $\quad\quad$ **else**
33: $\quad\quad\quad \mathbf{ID} \leftarrow \mathbf{ID} \cup \{ind\}$
34: $\quad\quad$ **end if**
35: $\quad$ **end if**
36: $\quad st_{i-1} \leftarrow P^{-1}(k_i, st_i)$
37: **end for**
38: send $\mathbf{ID}$ to client

Fig. 1: Pseudocode of Protocols in FAST

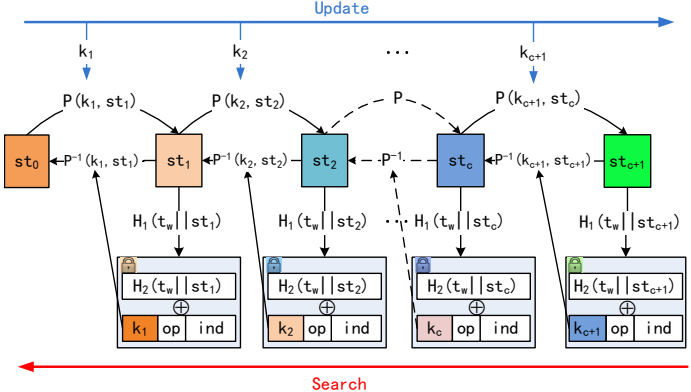

Fig. 2: Update and Search in FAST

of FAST, and has been optimized for I/O efficiency, which enables it to scale up to very large datasets.

### 5.1 I/O Deficiencies in FAST

As we discussed earlier, I/O often becomes a bottleneck of searchable encryption when dealing with large data, therefore it is necessary to optimize I/O for a searchable encryption scheme. Although FAST is optimal in terms of complexity and uses only symmetric primitives, it is poor in I/O efficiency.

Let us compare Sophos and FAST in term of I/O efficiency. The criteria are the server side index size, read efficiency and locality. In Sophos, the server side index size is $N \cdot l$ bits where $N$ is the number of document/keyword pairs in the index and $l$ is the identifier's size. This is optimal. To answer a search query with keyword $w$, Sophos needs to read $c \cdot l$ bits, where $c$ is the number of updates that contains $w$ since the initialization of the system. The read efficiency is 1, which means no unnecessary data is read and the read efficiency is optimal. The locality of Sophos is $c$ because the $c$ entries are located randomly in the encrypted index. In FAST, the server side index size is $N \cdot (l+1+\lambda)$ bits where $\lambda$ is the size of the random ephemeral key and the additional 1 bit is used to indicate the operation ("add" or "del"). To answer a search query with keyword $w$, FAST needs to read $c \cdot (l+1+\lambda)$ bits, which means the read efficiency is $(l+1+\lambda)/l$. The locality of searching a keyword in FAST is $c$, which is the same as Sophos.

As we can see, in FAST the server side index size and read efficiency are worse than those in Sophos. And the locality is as poor as Sophos. Take a closer look, we can find that the I/O deficiencies in FAST are related to its symmetric key construction. To use symmetric key primitives, FAST has to store the ephemeral key in the index. The key size $\lambda$ is often similar to $l$, which means a concrete overhead of 100% or so. However, if we do not store the key in the index, the server cannot re-generate previous states and cannot search. The question is: do we have to sacrifice I/O efficiency for computational efficient (i.e. using symmetric key primitives)?

### 5.2 FASTIO: the Idea

FASTIO is our answer to the above question. FASTIO still uses symmetric key primitives and maintains the same complexities and security level as FAST. However, it has a much better I/O efficiency. Concretely, the server side index size is at most $N \cdot (l+1)$ bits and the read efficiency is at most $(l+1)/l$. They are very close

```
Setup(λ, ⊥; ⊥)
    Client:
 1: k_s ←$ {0,1}^λ
 2: Σ ← empty map
    Server:
 3: T_e, T_c ← empty map

Update(k_s, Σ, ind, w, op; T_e)
    Client:
 4: (st, c) ← Σ[w]
 5: if (st, c) = ⊥ then
 6:     st ←$ {0,1}^λ
 7:     c ← 0
 8: end if
 9: u ← H_1(st||(c+1))
10: e ← (ind||op) ⊕ H_2(st||(c+1))
11: Σ[w] ← (st, c+1)
12: send (u, e) to server
    Server:
13: T_e[u] = e

Search(k_s, Σ, w; T_e, T_c)
    Client:
14: (st, c) ← Σ[w]
15: if (st, c) = ⊥ then
16:     return ⊥
17: end if
18: t_w ← F(k_s, h(w))
19: if c ≠ 0 then
20:     k_w ← st, st ←$ {0,1}^λ
21:     Σ[w] ← (st, 0)
22: else
23:     k_w ← ⊥
24: end if
25: send (t_w, k_w, c) to Server
    Server:
26: ID ← ∅
27: ID.add(T_c[t_w])
28: if k_w = ⊥ then
29:     return ID
30: end if
31: for i = 1 to c do
32:     u_i ← H_1(k_w||i)
33:     (ind, op) ← T_e[u_i] ⊕ H_2(k_w||i)
34:     if op = "del" then
35:         ID ← ID − {ind}
36:     else if op = "add" then
37:         ID ← ID ∪ {ind}
38:     end if
39:     delete T_e[u_i]
40: end for
41: T_c[t_w] ← ID
42: send ID to client
```

Fig. 3: Pseudocode of Protocols in FASTIO

to optimal. More importantly, in FASTIO we reduce the locality from $c$ to $\bar{c} + 1$, where $c$ is the number of updates that contains $w$ since *the initialization of the system* and $\bar{c}$ is the number of updates that contains $w$ since *the last search query*. The locality is optimal for forward private searchable encryption schemes if we also want to minimize the index size and read efficiency. As we will see later in the experiment section, the improvement is significant and FASTIO handles large datasets much better due to the optimal locality.

Our first key insight is that forward privacy does not require tracking all previous states. Indeed, forward privacy is defined with regard to new updates since the last search query. As long as the new updates are generated using a fresh state, forward privacy can be achieved. Our second key insight is that re-generating all states is not necessary for answering search queries. Think recursively: a search query $Q_i$ can be decomposed into two subqueries: (q1) find all matching documents that were updated before the last search query $Q_{i-1}$; (q2) find all matching documents that were updated after $Q_{i-1}$. It should be obvious that q1 is essentially $Q_{i-1}$. Therefore if the server keeps the result for $Q_{i-1}$, then when answering $Q_i$, it does not need to answer q1 again. Answering q2 requires only the state after $Q_{i-1}$, but not the whole history.

By storing the last search query result, we can improve I/O efficiency significantly. Recall that the poor index size and read efficiency in FAST is caused by storing the ephemeral keys. If the server does not need to re-generate states, the client does not need to store the keys. For locality, because the last search query result can be stored continuously and can be read altogether in one go, the overall locality can be reduced significantly. Storing the last search result does not affect security, because the server has already known the result. After each search, an adversarial server can always store the result even though the scheme does not require so. Therefore explicitly asking the server to store the result does not give the server any additional advantage.

### 5.3 The Construction

The protocols of FASTIO can be found in Fig. 3. In the Setup protocol, the client generates a long-term key for blinding keywords, and an empty map $\Sigma$ for storing states. The server generates two empty maps, $T_e$ is to be used to store the encrypted index and $T_c$ is to store the last search query results. For each keyword $w$, the client stores in $\Sigma$ two things: a state $st$ that is generated randomly at the first time $w$ is encountered or after each search query of $w$, and a counter $c$ to produce a sub-state for each update after a search query (until the next search query). Note that unlike in FAST that the state $st$ must evolve each time an update is made, in FASTIO $st$ stays unchanged in between two search queries even though there might be many updates. Although two updates may be based on the same $st$, in the Update protocol, the client can still make them unlinkable by hashing a sub-state that is the concatenation of $st$ and a counter (line 9 - 12). In the Search protocol, if there are some documents on the server side that contain the keyword $w$, the client generates $t_w$ (line 18). The purpose of $t_w$ is to enable the server to find the last search query result of $w$ in $T_c$. The client also retrieves the stored state as well as the counter, and checks whether they should be sent to the server as part of the search token (line 19 - 25). In this process, if the counter $c$ is non-zero, which means there have been updates since the last search query, then $st$ will be sent to the server and the client must generate a new random state $st$ and reset the counter at this point so that future updates can be made unlinkable. If the counter $c$ is zero, then there is no update since the last search query and the query result should be the same as the previous search. Therefore in this case $t_w$ alone is enough to retrieve the result and the client keeps $st$ secret. Search on the server side consists of two steps. The first step is to retrieve the last search query result using $t_w$ (line 27). Then the server may proceed to the second step (line 31 - 40). If there have been updates since the last search query, the server finds all updates using $k_w$ (the state) and the counter. Then the server finds the document identifier from each update and either removes it from or adds it to the result set depending on whether the update operation is del or add. The server also removes the update from $\mathbf{T}_e$ since its content has been revealed and is no long secret to the server. After all updates have been processed, the server returns the result set to the client and also stores it in $\mathbf{T}_c$ (line 41 - 42).

**Complexity Analysis** Complexity wise, FASTIO is almost the same as FAST, except that FASTIO has a better server side computational complexity for answering search queries. On the client side, the computational complexity for update and search are both $O(1)$. The client stores a $\lambda$-bit key and a map $\mathbf{\Sigma}$ whose size is $O(|\mathbf{W}|)$, where $|\mathbf{W}|$ is the total number of keywords. On the server side, the computational complexity for update is $O(1)$, and for search is $O(\bar{c})$ where $\bar{c}$ is the total number of updates that contains the keyword being searched since the last search query. The server stores a map $\mathbf{T}_e$ and $\mathbf{T}_c$ whose size in total is in the order of $O(N)$ where $N$ is the total number of file/keyword pairs. The communication complexity is $O(1)$ for update. For search, the communication complexity is $O(1)$ for the query and $O(|\mathsf{DB}(w)|)$ for the result set, which contains $|\mathsf{DB}(w)|$ matching files.

**Security Analysis** FASTIO achieves the same security level as FAST. We have the following theorem regarding the security of FASTIO:

***Theorem 2.*** Let $F$ be a pseudorandom function, $H_1$ and $H_2$ be two hash functions modeled as random oracles. Define leakage functions $\mathcal{L} = (\mathcal{L}_{\text{Setup}}, \mathcal{L}_{\text{Search}}, \mathcal{L}_{\text{Update}})$ as
$$\mathcal{L}_{\text{Setup}} = \bot$$
$$\mathcal{L}_{\text{Search}}(w) = (\mathsf{ap}(w), \mathsf{qp}(w))$$
$$\mathcal{L}_{\text{Update}}(\mathrm{i}, \mathrm{op}_i, w, \mathrm{ind}_i) = (i, \mathrm{op}_i, \mathrm{ind}_i)$$
Then FASTIO is an $\mathcal{L}$-adaptively-secure dynamic SSE with forward privacy

The proof can be found in Appendix B.

## 5.4 I/O Efficiency Analysis

FASTIO has a near optimal server side index size and read efficiency. In FASTIO, the server stores the index in two maps $\mathbf{T}_e$ and $\mathbf{T}_c$. In $\mathbf{T}_e$, each entry is $l+1$ bits where $l$ is the size of the document identifier. In $\mathbf{T}_c$, each entry is a set of document identifiers, thus the size is a multiple of $l$. In total, when there are $N$ document/keyword pairs, the size of the whole index is between $N \cdot l$ and $N \cdot (l+1)$ bits. To answer each search query, the server needs to read relevant entries in the index. The server needs to read in total $|\mathsf{DB}(w)|$ document identifier ($|\mathsf{DB}(w)|$ is the number of documents matching the keyword $w$). In a plaintext search, the server must read in at least all identifiers of the documents matching the keyword, which is in total $|\mathsf{DB}(w)| \cdot l$ bits. In FASTIO, some identifiers are read from $\mathbf{T}_e$ and some are read from $\mathbf{T}_c$. Entries in $\mathbf{T}_e$ adds 1 bit overhead per identifier for the operation type, and entries in $\mathbf{T}_c$ have no overhead. Therefore, the read efficiency of the search query is between 1 and $1 + \frac{1}{l}$. In practice, the document identifiers need to be long enough to be unique and $l$ is often large, e.g. 128. Then the server side index size and read efficiency in FASTIO are only less than 1% worse than optimal.

The locality of FASTIO is optimal under the constraint of forward privacy. As we have discussed in Section 3.5, locality and forward privacy are two irreconcilable notions. The implication is that, as Bost observed, the *worst-case* locality cannot be improved unless large modifications is done to the encrypted database. However, by utilizing the previous search query result, we can improve the *average-case* locality. In FASTIO, reading the previous search query result can be done in one go and if there have been $\bar{c}$ new updates after the last search query, additional $\bar{c}$ non-contiguous reads are needed to complete the search. Therefore the overall locality is $\bar{c} + 1$, which cannot be further improved without negatively affecting other properties.

## 6 PERFORMANCE EVALUATION

In this section, we evaluate the performance of FAST and FASTIO, and compare the results with Sophos, which is the most efficient forward private searchable encryption to date.

### 6.1 Implementation and Experiment Settings

We implemented FAST and FASTIO using C++. For Sophos, we use the C++ implementation[1] by the author. We use Crypto++ library[2] for the cryptographic operations: SHA256 for $H_1$ and $H_2$, and AES-128 for $F$ and $P$. To make the comparison fair, in our implementation, we use the same underlying libraries as in the Sophos implementation: Rocksdb[3] for storing key-value pairs and gRPC[4] for communication. The identifier length is set to 64-bit in all schemes. The server is deployed on an Alibaba Cloud ECS.i1.xlarge instance located in US West, which has 4 cores (Intel Xeon E5-2682v4, 2.5 GHz), 16GB RAM and $2 \times 104$ GB SSD disks. The client is deployed on a desktop PC located in China, which has 4 cores (Intel Core i5-3470, 3.7Ghz), 4 GB RAM and 500 GB hard disk.

### 6.2 Experiment Results

#### 6.2.1 Update Efficiency

We first show the performance of the update operation. As we explained earlier, the update operation in Sophos is based on a public key primitive, while update operations in FAST and FASTIO are all symmetric key based. We first measured the update efficiency of FAST, FASTIO and Sophos in a local setting in which we ran the client and the server on the same cloud instance. In this setting, the time we measured does not include latency caused by network, therefore it gives a better picture of the difference in computational efficiency. We then measured the update efficiency in a WAN setting, in which the server (in US) and the client (in China) are distributed. The results are shown in Table 1.

1. https://gitlab.com/sse/sophos
2. https://cryptopp.com
3. http://rocksdb.org
4. http://www.grpc.io

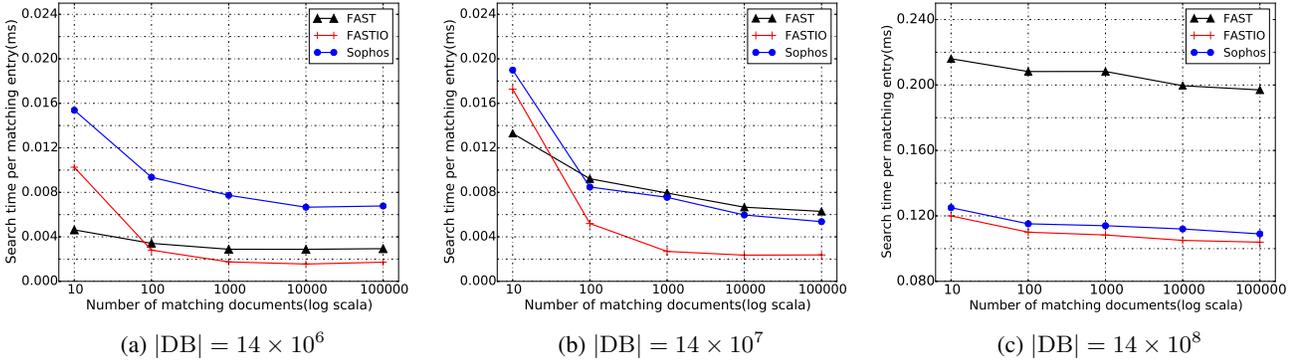

(a) $|DB| = 14 \times 10^6$
(b) $|DB| = 14 \times 10^7$
(c) $|DB| = 14 \times 10^8$

Fig. 4: Search time per matched document for FAST, FASTIO and Sophos.

As we can see, in the local setting, the throughput of FAST is about $11\times$, and FASTIO is about $15\times$, that of Sophos. In the WAN setting, the throughput of FAST is about $7\times$, and FASTIO about $10\times$, that of Sophos. The results confirm that FAST and FASTIO are superior in terms of update efficiency, compared to public key based Sophos.

|  |  | FAST | FASTIO | Sophos |
|---|---|---|---|---|
| Local | Throughput (ops/s) | 54060 | 76100 | 4890 |
|  | Single update time (ms) | 0.018 | 0.013 | 0.20 |
| WAN | Throughput (ops/s) | 21650 | 31080 | 2990 |
|  | Single update time (ms) | 0.046 | 0.032 | 0.334 |

TABLE 1: Update efficiency for FAST, FASTIO and Sophos

### 6.2.2 Search Efficiency – Index Processing

For FAST, FASTIO and Sophos, the search operation on the server side requires processing a list of encrypted index entries to find matching documents. The search efficiency thus depends crucially on the efficiency of processing the index. In Fig. 4, we show the performance of index processing in the three schemes.

We conducted three experiments with different database sizes $14 \times 10^6$, $14 \times 10^7$ and $14 \times 10^8$. In each experiment, we measure the total time on the server side (i.e. without counting network latency and token generation time on the client side) for searching keywords that have $10 - 10^5$ matching documents. We repeat 30 times and take the average, then divide it by the number of matching documents to get the time for processing a single entry. As we can see in the figure, the time for processing a single entry decreases when the number of entries increases. This is because there is a fixed cost for initializing the search, which is amortized into the per entry processing time. As the number of entries increases, the amortized initialization cost becomes less significant. When the database size is $14 \times 10^6$ and $14 \times 10^7$, FAST performs better than FASTIO and Sophos in the cases where the number of entries is small (10). This is because FAST has a smaller initialization cost, which is the result of an implementation level optimization.

We can observe the impact of I/O from Fig. 4. As we can see, for the same scheme and the same index size, the time we measured with the largest database ($14 \times 10^8$) is much higher than the time we measured with the smaller databases ($14 \times 10^6$ and $14 \times 10^7$). The performance degradation is about 2 orders of magnitude. The observation is in line with the results from previous studies (see Section 2). We can also see that FAST was impacted the most as it has the worst I/O efficiency.

In smaller databases where I/O does not dominates the processing time, FASTIO performs much better than Sophos. The performance difference is about 1 - 2 times. This is mainly because FASTIO uses only symmetric key operations. In large databases, the performance difference is not that large, but FASTIO still performs marginally better than Sophos.

### 6.2.3 Search Efficiency – Trace Simulation

The experiments in Section 6.2.2 do not fully reflect the search performance of FASTIO. In fact, it shows the *worst-case* performance of FASTIO. Recall that in FASTIO, the previous search results are stored to make search more efficient. In order to see how significant this improvement is, we also simulated the dynamic setting using traces. We generated 3 traces. Each trace is a list of update and search queries for a certain keyword. We fixed the length of the trace to 100,000. Each trace has a parameter $\alpha$ that is the probability of search queries, i.e. each query in the trace has a probability of $\alpha$ to be a search query and a probability of $1 - \alpha$ to be an update query. In the experiment, we let the client to replay the traces to simulate a real-world setting where updates and search queries are interleaved. We recorded the total time on the server side for each search query (from receiving search token to obtaining the results) in the trace.

In Fig. 5, we show the comparison of search efficiency with regard to the random traces. We used three different database sizes and three different query probabilities. In each sub-figure, the $x$-axis shows the sequence number of the query in the trace and the $y$-axis shows the search time. For FAST and Sophos, the search time increases almost monotonically in accordance to the number of update queries performed so far. Recall that in these two schemes, their indexes contain all entries from all previous updates. The index entries have to be stored in random locations in order to ensure forward privacy, this results in the decrease of I/O efficiency as the locality will increase monotonically. For FASTIO, as we can see, the search performance is much better, especially when search is more frequent. The difference is more significant for large databases. This is because in FASTIO, the index contains only entries since the last search query. The locality is much better than in the other two schemes.

In Fig. 6, we show the search time of FASTIO with regard to the traces. As we can see, despite the large number of updates, FASTIO's search time is always kept low. Obviously the more

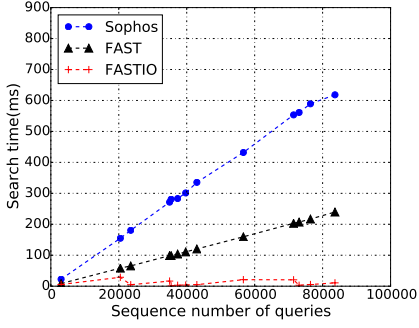
(a) $\alpha = 0.0001, |DB| = 14 \times 10^6$

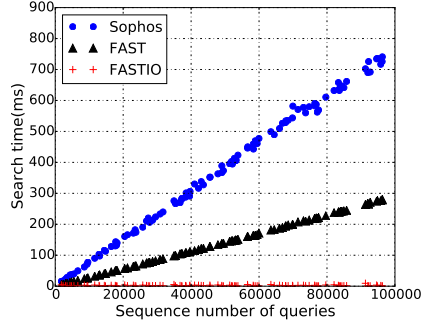
(b) $\alpha = 0.001, |DB| = 14 \times 10^6$

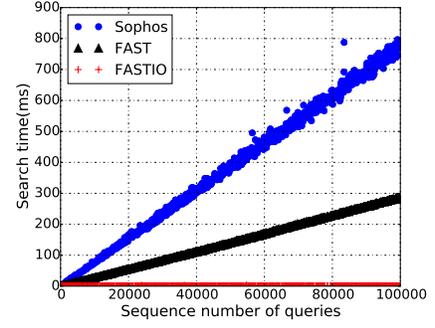
(c) $\alpha = 0.01, |DB| = 14 \times 10^6$

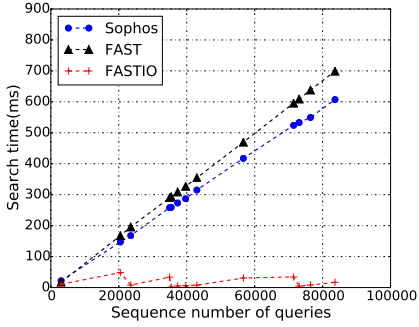
(d) $\alpha = 0.0001, |DB| = 14 \times 10^7$

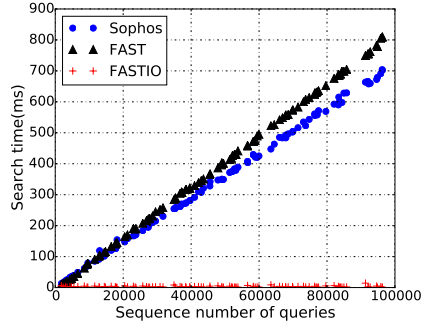
(e) $\alpha = 0.001, |DB| = 14 \times 10^7$

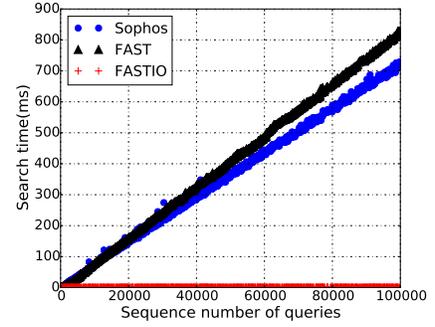
(f) $\alpha = 0.01, |DB| = 14 \times 10^7$

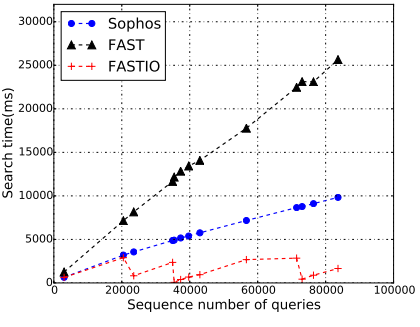
(g) $\alpha = 0.0001, |DB| = 14 \times 10^8$

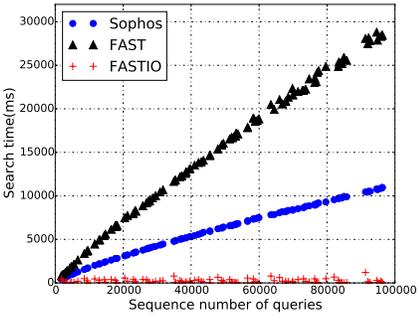
(h) $\alpha = 0.001, |DB| = 14 \times 10^8$

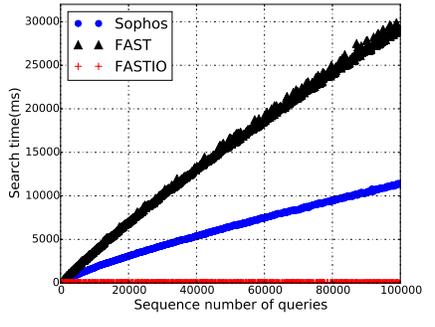
(i) $\alpha = 0.01, |DB| = 14 \times 10^8$

Fig. 5: Search efficiency by trace simulation.

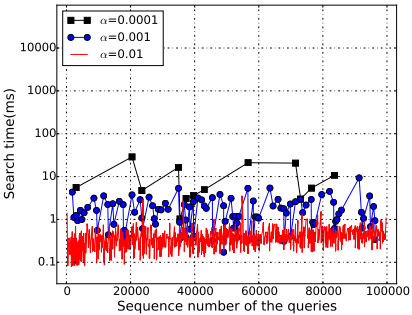
(a) $|DB| = 14 \times 10^6$

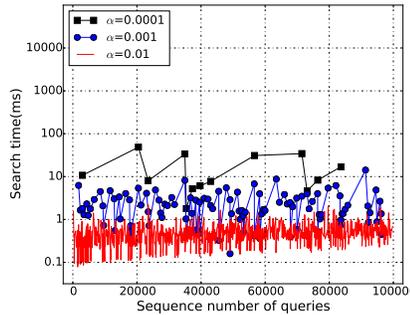
(b) $|DB| = 14 \times 10^7$

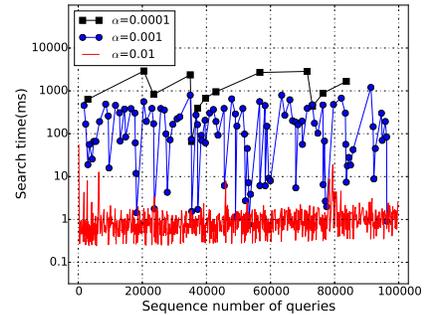
(c) $|DB| = 14 \times 10^8$

Fig. 6: Trace simulation for FASTIO. Search time is presented in log scale.

frequent the search queries are performed, the more performance gain we will see. This suggest that FASTIO has a better scalability, as frequent updates and large database size has a much less impact on its search performance than in the other two schemes.

# 7 CONCLUSION

Designing searchable symmetric encryption is not trivial if we want to combine efficiency and forward privacy, which are two irreconcilable properties. In this paper, we designed two forward private searchable symmetric encryption schemes, both achieve optimal computational and communicational complexity. As the first attempt, FAST utilized a state based approach and is computation friendly for its symmetric construction. Based on the idea from FAST and utilizing intrinsic leakage, our second construction FASTIO improves I/O efficiency significantly. Our experiment results demonstrated that our schemes are efficient and scalable.

**Future work and Open questions** For static searchable encryption, Cash et al. [10] gave theoretical bounds of locality, storage, and read efficiency and showed the relationship between them and the security notion. For dynamic searchable encryption, the relationship has not been well-understood and there has not been any theoretical study to establish the bounds. We would like to investigate in this direction so that we can understand better the limit and margin of further performance improvement.

In addition, modern data management system leverage distributed and parallel architecture to handle large amount of data. For SSE, search efficiency will sharply drop when the database is out of the capability of the single machine. In fact, even for plaintext database, it's not practical to be deployed on a single server. We would like to investigate searchable encryption in distributed and parallel settings.

# APPENDIX A
# PROOF OF THEOREM 1

*Proof:* We model the two hash functions $H_1$ and $H_2$ as random oracles. The oracles for $H_1$ and $H_2$ are identical except for different output lengths. Each oracle maintains a mapping $\mathbf{H}_i$ that stores input/output pairs $(in, out)$ where $in \in \{0,1\}^*$ and $out \in \{0,1\}^\ell$ where $\ell$ is the the length of the hash function's output. Given an input string $x$, the oracle first checks mapping $\mathbf{H}_i$, if there is an entry for $x$ then it returns the value associated with $x$ and terminates here. If there is not an entry for $x$ in $\mathbf{H}_i$, the oracle randomly picks a string $y$ from $\{0,1\}^\ell$, then stores $(x, y)$ in $\mathbf{H}_i$ and returns $y$.

We prove through a sequence of games. We start from $\mathbf{Real}_{\mathcal{A}}^\Pi(\lambda)$ and construct a sequence of games that differs slightly from the previous game and show they are indistinguishable. Eventually we reach the last game that is $\mathbf{Ideal}_{\mathcal{A},\mathcal{S}}^\Pi(\lambda)$. By the transitive property of the indistinguishability, we conclude that $\mathbf{Real}_{\mathcal{A}}^\Pi(\lambda)$ is indistinguishable from $\mathbf{Ideal}_{\mathcal{A},\mathcal{S}}^\Pi(\lambda)$ and complete our proof.

**Hybrid** $G_1$: $G_1$ is the same as $\mathbf{Real}_{\mathcal{A}}^\Pi(\lambda)$ except that instead of generating $t_w$ using $F$, the experiment maintain a mapping **Token** to store $(h(w), t_w)$ pairs. In the search protocol, when $t_w$ is needed, the experiment first checks whether there is an entry in **Token** for $h(w)$, if so returns the entry; otherwise randomly picks a $t_w$ in $\{0,1\}^l$ and stores the $(h(w), t_w)$ pair in **Token**. It's trivial to see that $G_1$ and $\mathbf{Real}_{\mathcal{A}}^\Pi(\lambda)$ are indistinguishable, otherwise we can distinguish a pseudo-random function $F$ and a truly random function.

**Hybrid** $G_2$: $G_2$ differs from $G_1$ in two places. Firstly, the experiment maintains a mapping **Key**. In the update protocol (line 9 in Fig. 1), after $k_{c+1}$ is generated, the experiment stores $(w||(c+1), k_{c+1})$ in **Key**. Secondly, instead of querying $H_1$ in the update protocol, a random string is used. The random string is used to program the random oracle when a search query is issued. More specially, Instead of doing $u \leftarrow H_1(t_w||st_{c+1})$, which is line 13 in the update protocol in Fig. 1, the experiment does the following:

$$u \xleftarrow{\$} \{0,1\}^\ell$$
$$\mathbf{L}[t_w||st_{c+1}] \leftarrow u$$

where $\mathbf{L}$ is a mapping maintained by the experiment. The second change is in the search protocol. The following is added after line 20 and line 21 of the search protocol:

**for** $i = 1$ **to** c **do**
 $\mathbf{H}_1[t_w||st_{c+1}] \leftarrow \mathbf{L}[t_w||st_{c+1}]$
 $k_i \leftarrow \mathbf{Key}[w||i]$
 $st_{i-1} \leftarrow P^{-1}(k_i, st_i)$
**end for**

where $\mathbf{H}_1$ is the table for random oracle $H_1$.

Now $G_1$ and $G_2$ behaves exactly the same except that in $G_2$, with some probability inconsistency in random oracle query results can be observed. In $G_2$, $\mathbf{H}_1$ is not updated immediately. For $t_w||st_{c+1}$, the corresponding value $u$ is generated in the update protocol. However, $u$ is only pushed into $\mathbf{H}_1$ when a search query is issued. After the update with state $t_w||st_{c+1}$, if the adversary queries $\mathbf{H}_1$ with $t_w||st_{c+1}$ before the next search query, it will get a value $u'$ such that with a overwhelming probability $u' \neq u$ because $\mathbf{H}_1[t_w||st_{c+1}]$ has not been updated and a random string $u'$ is chosen by the oracle in this case. If the adversary queries $\mathbf{H}_1$ with $t_w||st_{c+1}$ again after the next search query, $u$ will be updated to the $\mathbf{H}_1$ and the query result will be $u$. If the inconsistency is observed (we denote this event as **Bad**), the adversary knows it is in $G_2$. We have:

$$\Pr[G_1 = 1] - \Pr[G_2 = 1] \leq \Pr[\mathbf{Bad}]$$

Note that the event **Bad** can only happen if the adversary can query the oracle with $t_w||st_{c+1}$. Since $st_{c+1}$ is pseudorandom, the probability of the adversary choosing $st_{c+1}$ by chance is $2^{-\lambda} + \mathbf{negl}(\lambda)$. A PPT adversary can make at most $q_1 = \mathbf{poly}(\lambda)$ guesses, then $\Pr[\mathbf{Bad}] \leq \frac{q_1}{2^\lambda} + q_1 \cdot \mathbf{negl}(\lambda)$. The probability is negligible and $G_1$ and $G_2$ are indistinguishable.

**Hybrid** $G_3$: $G_3$ is obtained from $G_2$ in a similar way. In the update protocol, we replace the line $e \leftarrow (ind||op||k_{c+1}) \oplus H_2(t_w||st_{c+1})$ (line 12 in Fig. 1) with the following:

$$u \xleftarrow{\$} \{0,1\}^{l+\lambda+1}$$
$$\mathbf{E}[t_w||st_{c+1}] \leftarrow u$$
$$e \leftarrow (ind||op||k_{c+1}) \oplus u$$

where $\mathbf{E}$ is a mapping maintained by the experiment. Similarly, the following is added after line 20 and line 21 of the search protocol:

**for** $i = 1$ **to** c **do**
 $\mathbf{H}_2[t_w||st_{c+1}] \leftarrow \mathbf{E}[t_w||st_{c+1}]$
 $k_i \leftarrow \mathbf{Key}[w||i]$
 $st_{i-1} \leftarrow P^{-1}(k_i, st_i)$
**end for**

Using the same argument, we can conclude that $G_2$ and $G_3$ are indistinguishable.

**Hybrid** $G_4$: In $G_4$, the client side algorithms are changed as shown in Fig. 7. The main difference between $G_4$ and $G_3$ is that in $G_4$, $st$ and $k$ is generated on the fly when search is performed. In $G_4$ a mapping **Updates** is maintained by the experiment to record all the update requests since the last search query. Unlike in $G_3$ where the random oracle query results are chosen according to $t_w||st_{c+1}$, in $G_4$ the values are chosen randomly without the knowledge of $st$ (line 4 and 5, Fig. 7). Then when performing a search, $st_0$ is generated if this the first search query of $w$ (line 12 - 15, Fig. 7). Then $k_i, st_i$ are generated retrospectively and then the oracle is updated accordingly (line 21 - 31, Fig 7). The change however are not observable by the adversary. From the adversary's perspective, $G_3$ and $G_4$ behaves exactly the same: in both games the update protocol outputs two uniformly random strings, and the search protocol outputs $(t_w, k_w, c)$ that has the same distribution in the two games. Therefore they are perfectly indistinguishable.

$$\Pr[G_3 = 1] = \Pr[G_4 = 1]$$

$\mathbf{Ideal}_{\mathcal{A},\mathcal{S}}^\Pi(\lambda)$: in this game, a simulator must generate a view given only the leakage functions, but not the actual data and queries. For convenience, in the proof we will use *search pattern* and *update history* instead of using the search leakage function $\mathcal{L}_{\text{Search}}(w)$ directly. The two can be constructed from the leakage function. The search pattern $\mathsf{sp}(w) = \{j | q_j = w \text{ in } \mathsf{Hist}\}$ reveals which queries in the history are search queries with regard to $w$. The update history $\mathsf{up}(w) = \{(j, op_j, ind_j) | q_j = (ind_j, w, op_j) \text{ for each } q_j \text{ in } \mathsf{Hist}\}$ reveals which queries in the history are update queries with regard to $w$, as well as the update types and the document identifiers. We use $\mathsf{up}^{>k}(w)$ to denote the partial update history after the $k$-th query. It is clear that $\mathsf{sp}(w)$ can be obtained from $\mathsf{qp}(w)$ by throwing away all update requests, and $\mathsf{up}(w)$ can be obtained from $(\mathsf{ap}(w), \mathsf{qp}(w))$ by combining the information about update queries. They do not require additional information beyond what the leakage function allowed.

The simulator $\mathcal{S}$ is shown in Fig. 8. The simulator maintains two mappings for simulating random oracle queries, and a counter to record the number of updates since the initialization of the system. For each update query, two random strings are chosen. In

```
Setup()
  Client:
1: L, E ← empty map
2: v ← 0

Update(k_s, Σ, ind, w, op; T)
  Client:
3: Append (v + 1, op, ind) to Updates[w]
4: L[v + 1] ←$ {0,1}^ℓ
5: E[v + 1] ←$ {0,1}^{l+λ+1}
6: v ← v + 1
7: send (L[v], E[v]) to server

Search(k_s, Σ, w; T)
  Client:
8: if Token[h(w)] = ⊥ then
9:   Token[h(w)] ←$ {0,1}^λ
10: end if
11: t_w ← Token[h(w)]
12: if Σ[w] = ⊥ then
13:   Σ[w] ←$ {0,1}^λ
14: end if
15: st_0 ← Σ[w]
16: c = |Updates[w]|
17: [(u_0, op_0, ind_0), ..., (u_c, op_c, ind_c)] ← Updates[w]
18: if c = 0 then
19:   return ∅
20: end if
21: for i = 1 to c do
22:   if Key[w||i] ≠ ⊥ then
23:     k_i ← Key[w||i]
24:   else
25:     k_i ← {0,1}^λ
26:     Key[w||i] ← k_i
27:   end if
28:   st_i ← P(k_i, st_{i-1})
29:   H_1(t_w||st_i) ← L[u_i]
30:   H_2(t_w||st_i) ← E[u_i] ⊕ (ind_i||op_i||k_i)
31: end for
32: send (t_w, st_c, c) to Server
```

Fig. 7: Description of hybrid $G_4$

```
S.Setup()
  Client:
1: L, E ← empty map
2: v ← 0

S.Update()
  Client:
3: v ← v + 1
4: L[v] ←$ {0,1}^ℓ
5: E[v] ←$ {0,1}^{l+λ+1}
6: send (L[v], E[v]) to Server

S.Search(sp(w), uh(w))
  Client:
7: w̲ ← min sp(w)
8: if Token[w̲] = ⊥ then
9:   Token[w̲] ←$ {0,1}^λ
10: end if
11: t_w ← Token[w̲]
12: if Σ[w̲] = ⊥ then
13:   Σ[w̲] ←$ {0,1}^λ
14: end if
15: st_0 ← Σ[w̲]
16: c = |uh(w)|
17: [(u_0, op_0, ind_0), ..., (u_c, op_c, ind_c)] ← uh(w)
18: if c = 0 then
19:   return ∅
20: end if
21: for i = 1 to c do
22:   if Key[w̲||i] ≠ ⊥ then
23:     k_i ← Key[w̲||i]
24:   else
25:     k_i ←$ {0,1}^λ
26:     Key[w̲||i] ← k_i
27:   end if
28:   st_i ← P(k_i, st_{i-1})
29:   H_1(t_w||st_i) ← L[u_i]
30:   H_2(t_w||st_i) ← E[u_i] ⊕ (ind_i||op_i||k_i)
31: end for
32: send (t_w, st_c, c) to Server
```

Fig. 8: Description of simulator $S$

the search protocol, we use $\underline{w} = min\ \mathbf{sp}(w)$ to denote the very first index that $w$ appeared in search pattern. Each $\underline{w}$ uniquely identify an unknown keyword $w$ and the simulator can just use $\underline{w}$ without knowing $w$. The token $t_w$ is associated with $\underline{w}$. The simulator can use the update history to decide the update queries with regard to $w$ and their sequence number. Then states and keys can be generated given the update history. Then the random oracles are updated. The view produced by the simulator is perfectly indistinguishable from the one produced in $G_4$.

$$\Pr[\mathbf{Ideal}_{\mathcal{A},\mathcal{S}}^{\Pi}(\lambda) = 1] = \Pr[G_4 = 1]$$

Summing up, we have

$$\Pr[\mathbf{Real}_{\mathcal{A}}^{\Pi}(\lambda) = 1] - \Pr[\mathbf{Ideal}_{\mathcal{A},\mathcal{S}}^{\Pi}(\lambda) = 1] \leq \mathbf{negl}(\lambda)$$

Q.E.D. □

## APPENDIX B
## PROOF OF THEOREM 2

*Proof:* We model the two hash functions $H_1$ and $H_2$ as random oracles and prove through a sequence of games.

**Hybrid** $G_1$: $G_1$ is the same as $\mathbf{Real}_{\mathcal{A}}^{\Pi}(\lambda)$ except that instead of generating $t_w$ using $F$, the experiment maintain a mapping **Token** to store $(h(w), t_w)$ pairs. In the search protocol, when $t_w$ is needed, the experiment first checks whether there is an entry in **Token** for $h(w)$, if so returns the entry; otherwise randomly picks a $t_w$ in $\{0,1\}^l$ and stores the $(h(w), t_w)$ pair in **Token**. It's trivial to see that $G_1$ and $\mathbf{Real}_{\mathcal{A}}^{\Pi}(\lambda)$ are indistinguishable, otherwise we can distinguish a pseudo-random function $F$ and a truly random function.

**Hybrid $G_2$**: $G_2$ differs from $G_1$ in that instead of querying $H_1$ in the update protocol, a random string is used. The random string is used to program the random oracle when a search query is issued. More specially, Instead of doing $u \leftarrow H_1(st||(c+1))$, which is line 9 in the update protocol in Fig. 3, the experiment does the following:

$$u \xleftarrow{\$} \{0,1\}^\ell$$
$$\mathbf{UT}[st||(c+1)] \leftarrow u$$

where $\mathbf{UT}$ is a mapping maintained by the experiment. The second change is in the search protocol. The following is added in between line 19 and line 20 of the search protocol:

    **for** $i = 1$ **to** c **do**
        $\mathbf{H}_1[st||i] \leftarrow \mathbf{UT}[st||i]$
    **end for**

where $\mathbf{H}_1$ is the table for random oracle $H_1$.

Now $G_1$ and $G_2$ behaves exactly the same except that in $G_2$, with some probability inconsistency in random oracle query results can be observed. In $G_2$, $\mathbf{H}_1$ is not updated immediately. For $st||(c+1)$, the corresponding value $u$ is generated in the update protocol. However, $u$ is only pushed into $\mathbf{H}_1$ when a search query is issued. After the update with state $st$ and counter $c+1$, if the adversary queries $\mathbf{H}_1$ with $st||(c+1)$ before the next search query, it will get a value $u'$ such that with a overwhelming probability $u' \neq u$ because $\mathbf{H}_1[st||(c+1)]$ has not been updated and a random string $u'$ is chosen by the oracle in this case. If the adversary queries $\mathbf{H}_1$ with $st||(c+1)$ again after the next search query, $u$ will be updated to the $\mathbf{H}_1$ and the query result will be $u$. If the inconsistency is observed (we denote this event as **Bad**), the adversary knows it is in $G_2$. We have:

$$\Pr[G_1 = 1] - \Pr[G_2 = 1] \leq \Pr[\mathbf{Bad}]$$

Note that the event **Bad** can only happen if the adversary can query the oracle with $st||(c+1)$. Since $st \xleftarrow{\$} \{0,1\}^\lambda$, the probability of the adversary choosing $st$ by chance is $2^{-\lambda}$. A PPT adversary can make at most $q_1 = \mathbf{poly}(\lambda)$ guesses, then $\Pr[\mathbf{Bad}] \leq \frac{q_1}{2^\lambda}$. The probability is negligible and $G_1$ and $G_2$ are indistinguishable.

**Hybrid $G_3$**: $G_3$ is obtained from $G_2$ in a similar way. In the update protocol, we replace the line $e \leftarrow (ind||op) \oplus H_2(st||(c+1))$ with the following:

$$u \xleftarrow{\$} \{0,1\}^{l+1}$$
$$\mathbf{E}[st||(c+1)] \leftarrow u$$
$$e \leftarrow (ind||op) \oplus u$$

where $\mathbf{E}$ is a mapping maintained by the experiment. Similarly, the following is added in after line 19 of the search protocol:

    **for** $i = 1$ **to** c **do**
        $\mathbf{H}_2[st||i] \leftarrow \mathbf{E}[st||i]$
    **end for**

Using the same argument, we can conclude that $G_2$ and $G_3$ are indistinguishable.

**Hybrid $G_4$**: In $G_4$, the client side protocols are changed as shown in Fig. 9. The main difference between $G_4$ and $G_3$ is that in $G_4$, $st$ is generated on the fly when search is performed. In $G_4$ a mapping **Updates** is maintained by the experiment to record all the update requests since the last search query. Unlike in $G_3$ where the random oracle query results are chosen according to $st||(c+1)$,

---

Setup()

  *Client:*
1: $\mathbf{UT}, \mathbf{E} \leftarrow$ empty map
2: $v \leftarrow 0$

Update($k_s, \Sigma, ind, w, op; \mathbf{T}_e$)

  *Client:*
1: Append $(v+1, op, ind)$ to **Updates**$[w]$
2: $\mathbf{UT}[v+1] \xleftarrow{\$} \{0,1\}^\ell$
3: $\mathbf{E}[v+1] \xleftarrow{\$} \{0,1\}^{l+1}$
4: $v \leftarrow v + 1$
5: send ($\mathbf{UT}[v], \mathbf{E}[v]$) to server

Search($k_s, \Sigma, w; \mathbf{T}_e, \mathbf{T}_c$)

  *Client:*
6: **if** **Token**$[h(w)] = \perp$ **then**
7:     **Token**$[h(w)] \xleftarrow{\$} \{0,1\}^\lambda$
8: **end if**
9: $t_w \leftarrow$ **Token**$[h(w)]$
10: **if** **Updates**$[w] = \perp$ **then**
11:     $k_w \leftarrow \perp$
12: **else**
13:     $k_w \xleftarrow{\$} \{0,1\}^\lambda$
14:     $c = |\mathbf{Updates}[w]|$
15:     $\{(v_1, op_1, ind_1),...,(v_c, op_c, ind_c)\} \leftarrow$ **Updates**$[w]$
16:     **for** $1 \leq i \leq c$ **do**
17:         $\mathbf{H}_1(k_w||i) \leftarrow \mathbf{UT}[v_i]$
18:         $\mathbf{H}_2(k_w||i) \leftarrow \mathbf{E}[v_i] \oplus (op_i||ind_i)$
19:     **end for**
20: **end if**
21: send $(t_w, k_w, c)$ to Server
22: **Updates**$[w] \leftarrow \perp$

Fig. 9: Description of hybrid game $G_4$

in $G_4$ the values are chosen randomly without the knowledge of $st$ (line 1 and 2, Fig 9). Then when performing a search, $k_w$ (i.e. $st$) is generated retrospectively and then the oracle is updated accordingly (line 13 - 19, Fig 9). The change however are not observable by the adversary. From the adversary's perspective, $G_3$ and $G_4$ behaves exactly the same: in both games the update protocol outputs two uniformly random strings, and the search protocol outputs $(t_w, k_w, c)$ that has the same distribution in the two games. Therefore they are perfectly indistinguishable

$$\Pr[G_3 = 1] = \Pr[G_4 = 1]$$

**Ideal**$_{\mathcal{A},\mathcal{S}}^{\Pi}(\lambda)$: in this game, a simulator must generate a view given only the leakage functions, but not the actual data and queries. For convenience, in the proof we will use *search pattern* and *update history* instead of using the search leakage function $\mathcal{L}_{\text{Search}}(w)$ directly. The two can be constructed from the leakage function. The search pattern $\mathsf{sp}(w) = \{j | q_j = w \text{ in Hist}\}$ reveals which queries in the history are search queries with regard to $w$. The update history $\mathsf{up}(w) = \{(j, op_j, ind_j) | q_j = (ind_j, w, op_j) \text{ for each } q_j \text{ in Hist}\}$ reveals which queries in the history are update queries with regard to $w$, as well as the update

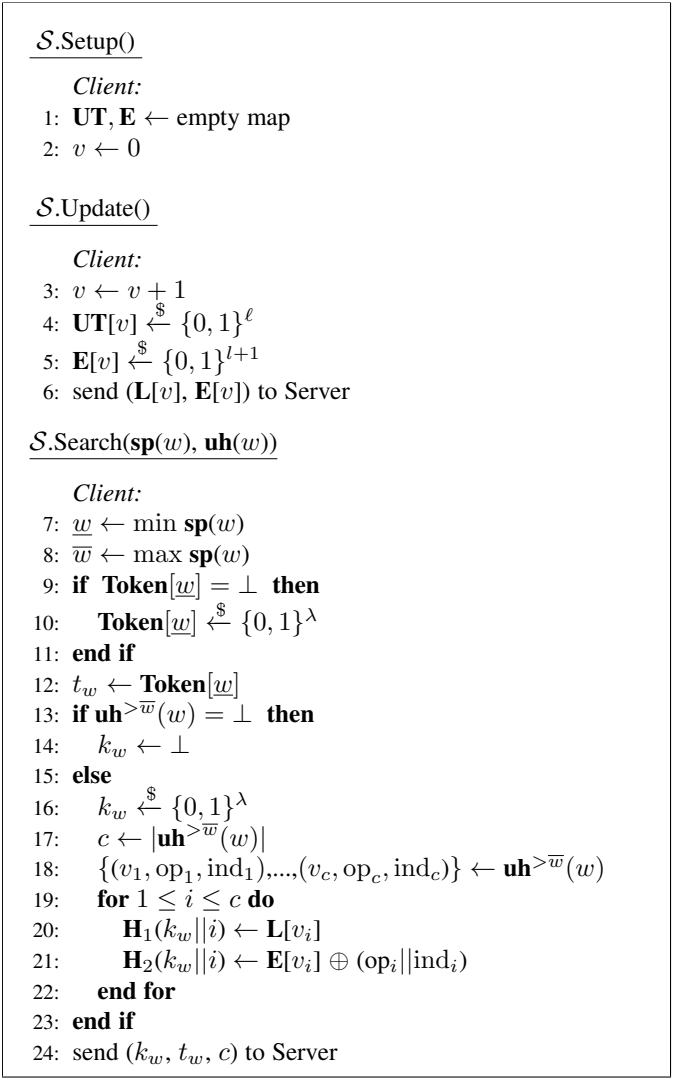

Fig. 10: Description of simulator $\mathcal{S}$

indistinguishable from the one produced in $G_4$.

$$\Pr[\textbf{Ideal}_{\mathcal{A},\mathcal{S}}^{\Pi}(\lambda)=1] = \Pr[G_4=1]$$

Summing up, we have

$$\Pr[\textbf{Real}_{\mathcal{A}}^{\Pi}(\lambda)=1] - \Pr[\textbf{Ideal}_{\mathcal{A},\mathcal{S}}^{\Pi}(\lambda)=1] \leq \textbf{negl}(\lambda)$$

Q.E.D. □

types and the document identifiers. We use $\mathsf{up}^{>k}(w)$ to denote the partial update history after the $k$-th query. It is clear that $\mathsf{sp}(w)$ can be obtained from $\mathsf{qp}(w)$ by throwing away all update requests, and $\mathsf{up}(w)$ can be obtained from $(\mathsf{ap}(w), \mathsf{qp}(w))$ by combining the information about update queries. They do not require additional information beyond what the leakage function allowed.

The simulator $\mathcal{S}$ is shown in Fig. 10. The simulator maintains two mappings for simulating random oracle queries, and a counter to record the number of updates since the initialization of the system. For each update query, two random strings are chosen. In the search protocol, we use $\underline{w} = min\ \mathsf{sp}(w)$ to denote the very first index that $w$ appeared in search pattern and $\overline{w} = max\ \mathsf{sp}(w)$ to denote the last index that $w$ been searched. Each $\underline{w}$ uniquely identify an unknown keyword $w$ and the simulator can just use $\underline{w}$ without knowing $w$. The token $t_w$ is associated with $\underline{w}$. The simulator can use the update history to decide whether there have been new updates since the last search query. If so, a random $k_w$ is generated and the number of updates after the last search query can be decided given the update history. Then the random oracles are updated. The view produced by the simulator is perfectly